# Physical design of FEL injector based on performance-enhanced EC-ITC RF gun

HU Tong-ning(胡桐宁)[1], CHEN Qu-shan(陈曲珊)[1], PEI Yuan-ji(裴元吉)[2; 1)], LI Ji(李骥)[2], QIN Bin(秦斌)[1]

[1] *State Key Laboratory of Advanced Electromagnetic Engineering and Technology, Huazhong University of Science and Technology, Wuhan 430074, China;*

[2] *National Synchrotron Radiation Laboratory, University of Science and Technology of China, Hefei 230029, China*

**Abstract:** To meet requirements of high performance THz-FEL (Free Electron Laser), a compact scheme of FEL injector was proposed. Thermionic cathode was chosen to emit electrons instead of photo-cathode with complex structure and high cost. The effective bunch charge was improved to ~200pC by adopting enhanced EC-ITC (External Cathode Independently Tunable Cells) RF gun to extract micro-bunches, and back bombardment effects were almost eliminated as well. Constant gradient accelerator structures were designed to improve energy to ~14MeV, while focusing system was applied for emittance suppressing and bunch state maintenance. Physical design and beam dynamics of key components for FEL injector were analyzed. Furthermore, start-to-end simulations with multi-pulses were performed by using homemade MATLAB and Parmela. The results show that continual high brightness electron bunches with low energy spread and emittance could be obtained stably.

## 1  Introduction

Nowadays, high quality electron beam sources have been focused due to rapid development of FEL facilities. As a key component, RF gun has been used widely because of its high electric field strength. Comparing with expensive and complicated photocathode RF gun, thermionic RF gun with compact and simple structure generates short bunches relying on self-bunching effect of electrons in RF standing-wave fields. According to researches of many institutions recently, thermionic RF gun still has potentials of generating high brightness bunches and compressing bunch length to have sufficient peak current to drive FEL [1,2].

In order to take advantages of structure compactness and high performance potentials of thermionic RF gun, Tsinghua University (THU) and China Academy of Engineering Physics (CAEP) have got outstanding achievements by shortening the first cavity and adjusting electric field strength ratio of multi-cavities[3,4]. Ref. [5,6,7] proposed preliminary concept of ITC RF gun with much simpler structure, which has two independent cavities without $\alpha$-magnets, while feed-in powers and phases could be adjusted independently. An improvement method by using external cathode instead of embed one has been performed in National Synchrotron Radiation Laboratory (NSRL), which almost eliminated back bombardment effect by reducing electric field influences on cathode surface, and effective bunch charge has been increased from tens of pC to ~130pC within 4.5ps FWHM (Full Width at Half Maximum) length, while energy spread (FWHM) is only ~0.2% [8,9,10]. However, main performances, especially effective bunch charge, are hard to be increased more due to space charge effect of high current electron beam.

To meet strict requirements of high performance THz-FEL in Huazhong University of Science and Technology (HUST), the effective bunch charge within 5ps length (FWHM) must be increased to ~200pC with low energy spread and low transverse emittance [11,12]. By adjusting structure and RF parameters, properties of EC-ITC RF gun have been enhanced. Combining with equal gradient travelling-wave accelerator and focusing system consists of short magnetic lens and solenoids, a design scheme of compact high brightness FEL injector is proposed and shown as Fig. 1, which is mainly comprised of EC-ITC RF gun, traveling-wave accelerator and focusing system, while specific design targets are listed in Table 1.

1).Corresponding author (email: yjpei@ustc.edu.cn).

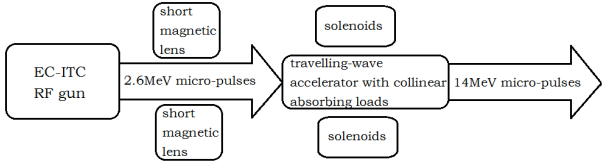

Fig. 1　Schematic of FEL injector.

Table. 1　Specific targets of FEL injector.

| Parameter | Value |
| --- | --- |
| Beam energy | ~14MeV |
| Micro-pulse width (FWHM) | ~5ps |
| Micro-pulse charge | 200~300pC |
| Energy spread (rms) | 0.5% |
| Transverse normalized emittance (rms) | 6.5 $\pi$mm·mrad |
| Micro-pulse repetition rate | 2856MHz |
| Macro-pulse width | 2~6us |
| Macro-pulse repetition rate | 10~50Hz |

To obtain high quality bunches described above, components must cooperate with each other closely. Short high brightness micro-pulses are bunched from DC beam provided by grided electron gun, and extracted by ITC standing-wave cavities, then enters into travelling-wave accelerator for ~14MeV energy gain, while attached short magnetic lens and solenoids are applied to compress bunch size further more.

## 2　Performance-enhanced EC-ITC RF gun

Considering beam quality of the micro-pulse might be deteriorated when it transmits from EC-ITC RF gun to drift tube and travelling-wave accelerator due to space charge effect of high current beam, it's necessary to improve design targets of RF gun to guarantee high quality extracted bunches. As mentioned in last section, RF gun structures adopted by reference [8,9,10] could generate short bunches with low energy spread and low transverse emittance, but the effective bunch charge was too low to drive FEL. To solve this issue, grided electron gun with double anodes is preferred to provide DC beam instead of diode electron gun, standing-wave cavity dimensions are optimized, and RF parameters are tuned independently as well. As a result, EC-ITC RF gun performances are enhanced substantially, and Fig. 2 demonstrates the schematic of structures.

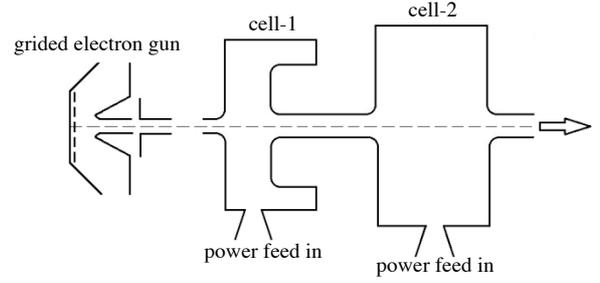

Fig. 2　Schematic of EC-ITC RF gun.

Cathode-grid assembly with model number Y646 produced by EIMAC Corp. is chosen to emit electrons, while double anodes are adopted to lengthen gunshot range to ~40mm and compress waist radius to ~1mm. By adjusting dimensions and anode voltages using Poisson and Parmela[13,14] codes with ring-based algorithms, beam envelope shown in Fig. 3 is obtained, with 15keV kinetic energy, 5A/cm$^2$ current intensity and 6$\pi$mm·mrad transverse emittance (rms).

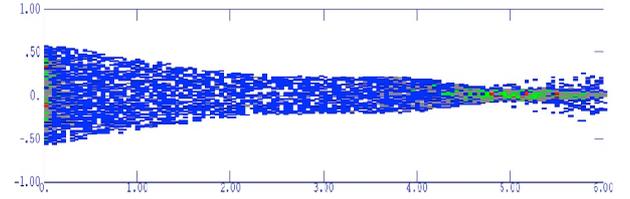

Fig. 3　Beam envelope extracted from grided electron gun.

DC beam extracted from grided electron gun enters into ITC cavities. Over 45% particles are caught by the fist cavity and bunched into 3ps length (FWHM), while the second cavity increases bunch energy to 2.6MeV and compresses length even more. Maximum electric field strength on axis of cavities provided by Superfish is illustrated in Fig. 4, and output bunch states of both cells calculated by Parmela are shown in Fig. 5, with beam dynamic results listed in Table 2 in details.

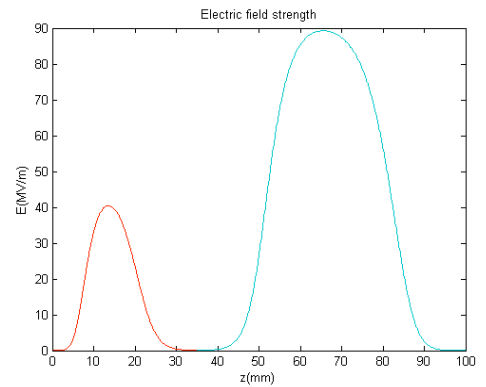

Fig. 4　Electric field strength on beam axis of EC-ITC cavities.

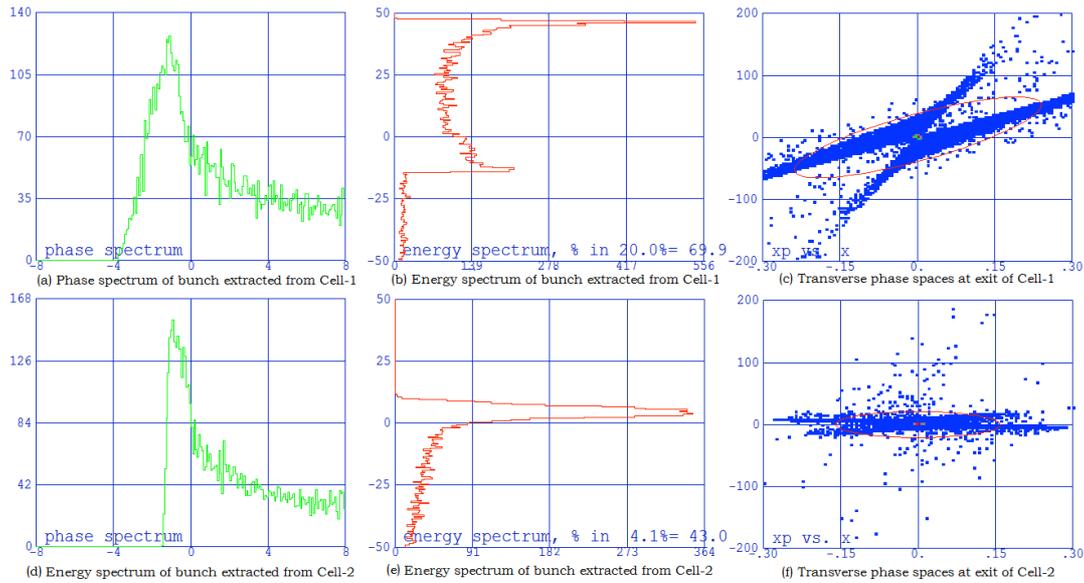

Fig. 5 Output bunch states of EC-ITC RF gun.

Table. 2 Output bunch properties of EC-ITC RF gun.

| Parameter | Target |
| --- | --- |
| Bunch energy | 2.6 MeV |
| Micro-pulse width (FWHM) | ~1.5 ps |
| Micro-pulse effective charge | 201pC |
| Energy spread (FWHM) | 0.27% |
| Energy spread (rms) | 0.28% |
| Transverse emittance (rms) | 6.5 $\pi$mm·mrad |
| Macro-pulse current | 0.574A |
| Bunch radius | 2.8 mm |

Obviously, the performance-enhanced EC-ITC RF gun has transferred DC beam into micro-pulses with 1.5ps width (FWHM) and improved effective bunch charge to ~200pC. Furthermore, both energy spread and transverse emittance are small enough to meet overall targets as well.

## 3  Focusing system

Main properties of pulsed short bunches with below 3MeV energy and over 100A current intensity, such as length, radius and transverse emittance, will deteriorate in the transmitting progress due to space charge effect. For the sake of bunch maintenance and improvement, focusing magnetic field is designed generally to attach with drift tube and travelling-wave accelerator, shown as Fig. 6. Taking into account of compactness, focusing system consists of short magnetic lens and solenoids. The former is out of drift tube between EC-ITC RF gun and travelling-wave accelerator, that high magnetic strengths could be obtained in the center while the attenuations of sideward's are fast enough to avoid affecting the obits of low energy electrons near cathode. In addition, solenoids composed with current tunable coils could restrain weak space charge force and radial defocusing in accelerator to make sure thigh brightness bunches would be extracted stably.

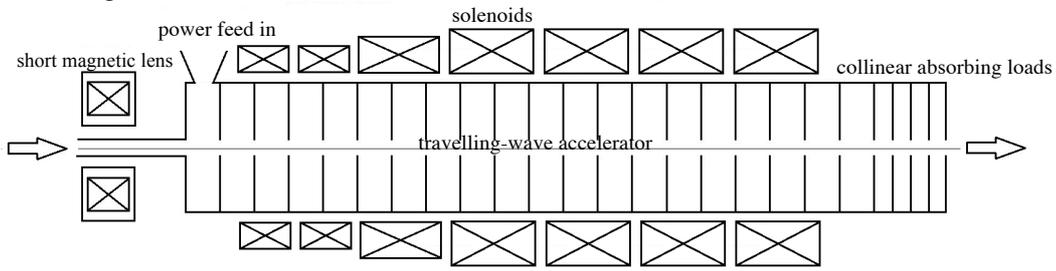

Fig. 6 Focusing system and LINAC.

Magnetic field distributions of the focusing system near beam axis designed by Poisson are given by Fig. 7. Beam dynamic results indicate that FEL injector will generate micro-pulses with small stable dimensions when peak magnetic field strengths of short magnetic lens and solenoids reach 1600Gauss and 2000Gauss, respectively.

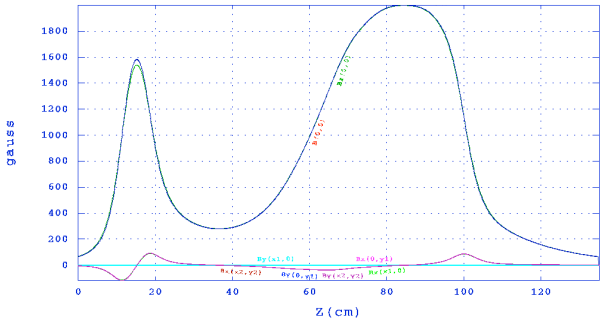

Fig. 7　Magnetic field generated by focusing system.

## 4　Travelling-wave accelerator

To make micro-pulse energy reach ~14MeV, travelling-wave accelerator consists of one power fed-in cavity, 19 normal accelerating cells with $2\pi/3$ mode, and 4 collinear absorbing loads, is designed. By coating wave-absorbing materials on cavity inner surfaces, collinear absorbing loads are used to avoid asymmetric fields caused by single coupling cavity, and reduce the main focusing magnet size to realize structure compactness.

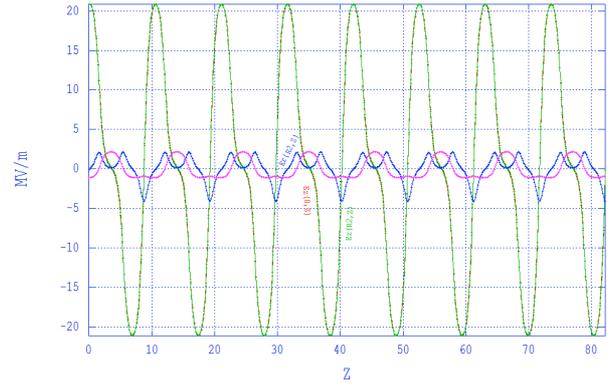

Fig. 8　Electro-magnetic field distributions near axis of LINAC.

After entering into accelerator, electron beam could gain energy from travelling-wave fields illustrated in Fig. 8, maximum gain could be obtained by tuning accelerating phases. Bunch property variations both with and without space charge force in drift tube and accelerator calculated by Parmela are shown in Fig. 9. Apparently, energy gain curves are the same. However, comparing to slight fluctuations of energy spread and transverse emittance without space charge force, the calculating results considering space charge effect deteriorated first, then tend to stable in the latter half of accelerator because space charge effect could be ignored when bunch gets close to relativistic velocity.

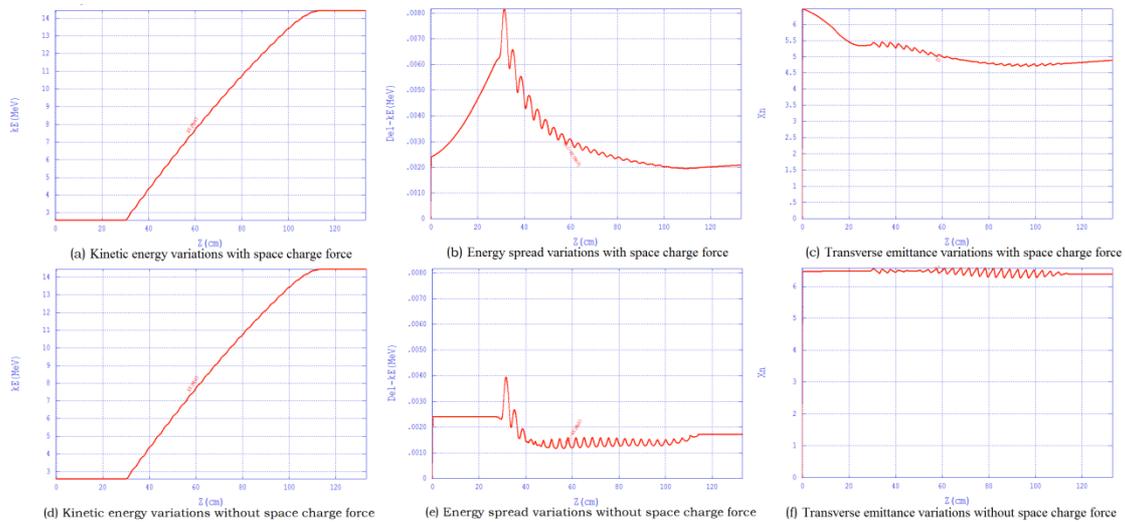

Fig. 9　Bunch property variations in drift tube and LINAC.

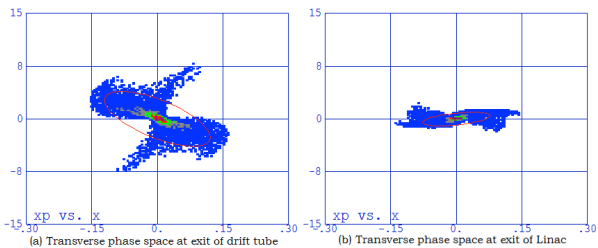

Fig. 10　Transverse phase spaces at exits of drift tube and LINAC.

For comparison, transverse phase spaces at the ends of drift tube and LINAC are given by Fig. 10. Micro-pulse states extracted from travelling-wave accelerator are shown in Fig. 11, while Table 4 gives specific performances. The results show that main parameters such as energy spread and transverse emittance are better than before, except bunch has been lengthen to ~4ps.

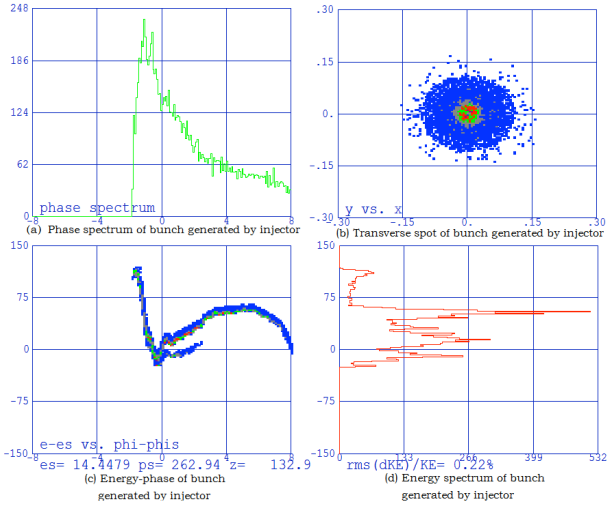

Fig. 11 Output bunch states of LINAC with space charge force.

Table. 4  Output bunch properties of LINAC.

| Parameter | Target |
|---|---|
| Bunch energy | 14.4MeV |
| Micro-pulse width (FWHM) | ~4 ps |
| Micro-pulse effective charge | 201 pC |
| Energy spread (rms) | 0.22% |
| Transverse emittance (rms) | 4.8 $\pi$mm·mrad |
| Macro-pulse current | 0.574 A |
| Bunch spot radius | 0.9 mm |

## 5  Multi-pulse simulations

For the sake of performance stability of THz-FEL, it's vital for FEL injector to provide high quality micro-pulses continually. Since explicit properties can't be given directly by Parmela with only one reference particle used, a home-made MATLAB code based on it has been written to analyze multi-pulse situation, which computes space charge force by Lorentz-transforming the particles positions and field maps into the average rest frame of the beam forces between electrons in rest frame, then applies static forces to various rings of the cylindrical map assuming a constant charge density inside a ring[15].

DC beams of three RF periods from grided electron gun enter into ITC cavities and following accelerator continually, then micro-pulses are extracted by FEL injector, with phase and energy spectra shown as Fig. 12.

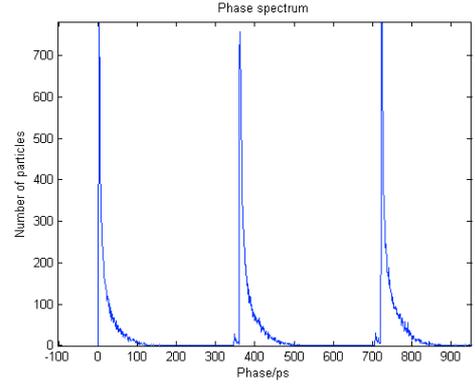

(a) Phase spectra of 3 micro-pulses.

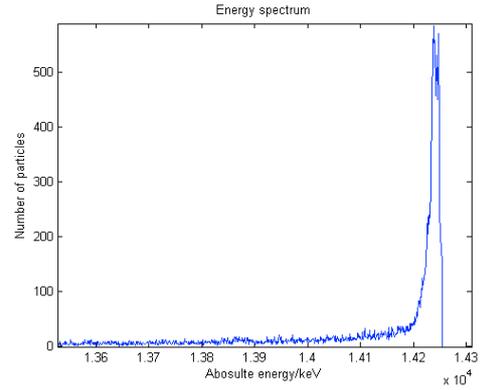

(b) Energy spectra of 3 micro-pulses.

Fig. 12  Phase and energy spectra of 3 micro-pulses with continual DC beam input.

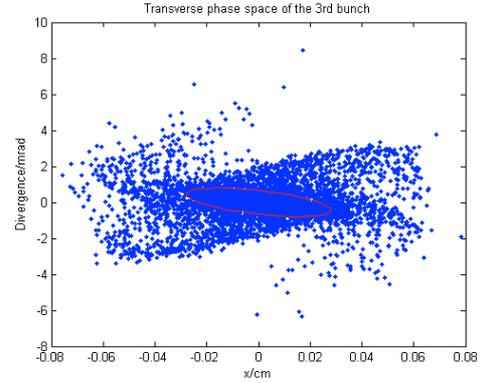

Fig. 13  Transverse phase space of the 3rd bunch.

Table. 5  Main properties of each bunch with 3-pulses input.

| Parameter | The 1st bunch | The 2nd bunch | The 3rd bunch |
|---|---|---|---|
| Effective electric charge of bunch head | 210pC | 228pC | 233pC |
| Energy spread (rms) of effective bunch head | 0.11% | 0.08% | 0.09% |
| Transverse normalized emittance (rms) | 10.8 mm·mrad | 9.0 mm·mrad | 5.0 mm·mrad |
| Bunch length(FWHM) | 3.5ps | 5ps | 3ps |

Table 5 gives main properties of each bunch separately. Apparently, all of them meet THz-FEL requirements. Furthermore, because of longitudinal space charge effect and influences of former beam tails, the third bunch with transverse phase space shown as Fig. 13, has got stable properties and should be used to measure injector performances. In short, continual stable high quality micro-pulses are obtained by using this type of FEL injector.

## 6  Conclusion

Performance-enhanced EC-ITC RF gun has many outstanding merits such as simple structure, low cost and ability of generating high brightness pulsed bunches from DC beam by adjusting fed-in powers and phases independently. The effective bunch charge is over 200pC within ~1.5ps micro-pulse width (FWHM), while energy spread (FWHM) is only ~0.3% and transverse emittance (rms) is less than 7 $\pi$mm·mrad . The following travelling-wave accelerator with collinear absorbing loads could improve bunch energy to ~14MeV and compress bunch spot radius to 0.9mm, combining with focusing system. In the latter half of accelerator, bunches get close to light velocity that space charge force could be ignored and each property of micro-pulses maintain stable with 0.22% energy spread(rms) and 4.8 $\pi$mm·mrad transverse emittance (rms). Consequently, analysis results indicate that high quality micro-pulses suitable for THz-FEL could be generated by this type of FEL injector stably.